\documentstyle[aps,epsf,twocolumn]{revtex}
\begin{document}
\title{Magnetic Field Induced Charging Effects in Josephson Junction Arrays}
\author{Sergei Sergeenkov\cite{byline}}
\address{Instituto de Fis\'ica, Universidade Federal do Rio Grande do Sul,
91501-970 Porto Alegre, Brazil}
\date{\today}
\address{~}
\address{
\centering{
\begin{minipage}{16cm}
A magnetic field induced electric polarization
and the corresponding change of an effective
junction capacitance are considered within a $3D$ model of disordered
Josephson junction arrays. At some threshold field (near the Josephson
network critical field), the effective junction charge and the related
capacitance are shown to reach a maximum and to change a sign, respectively.
A possibility to observe the predicted effects in artificially prepared
arrays of superconducting grains is discussed.
\end{minipage}
}}
\maketitle
\narrowtext
Recently, a variety of different field-induced phenomena in
high-$T_c$ superconductors (HTS), related to their extrinsic or intrinsic
granularity, have been reported (see,
e.g.,~\cite{ref1,ref2,ref3,ref4,ref5,ref6} and further references
therein). In particular, an appearance of an electric-field induced non-zero
magnetization signal in granular superconductors (due to the
Dzyaloshinski-Moria type interaction between an applied electric field and
an effective magnetic field of circulating Josephson currents) was
predicted~\cite{ref6}. At the same time, an artificially prepared islands
of superconducting grains, well-described by the various models of Josephson
junction arrays (JJA), proved useful in studying (both theoretically and
experimentally) the charging effects in these systems, ranging from Coulomb
blockade of Cooper pair tunneling and Bloch oscillations~\cite{ref7,ref8}
to propagation of quantum ballistic vortices~\cite{ref9}.

The present paper addresses yet another related phenomenon which is
actually dual to the above-mentioned analog of magnetoelectric
effect (described in Ref.~\cite{ref6}). Specifically, we discuss a possible
appearance of a non-zero
electric polarization and the related change of the charge balance in the
system of weakly-coupled superconducting junctions (modelled by the random
$3D$ JJAs) under the influence of an applied magnetic field.

To achieve our goal, we apply the so-called model of disordered $3D$ JJAs
based on the well-known tunneling Hamiltonian
(see, e.g.,~\cite{ref10})
\begin{equation}
{\cal H}=\sum_{ij}^NJ_{ij}[1-\cos \phi _{ij}(\vec H)],
\end{equation}
where
\begin{equation}
\phi_{ij}(\vec H)=\phi _{ij}(0)-A_{ij}(\vec H),
\end{equation}
with
\begin{equation}
\phi _{ij}(0)=\phi _i-\phi _j,
\end{equation}
and
\begin{equation}
A_{ij}(\vec H)=\frac{\pi}{\Phi _0}(\vec H\times \vec R_{ij})\vec r_{ij},
\end{equation}
where
\begin{equation}
\vec r_{ij}=\vec r_i-\vec r_j, \qquad \vec R_{ij}=(\vec r_i+\vec r_j)/2,
\end{equation}
which describes an interaction between $N$ superconducting grains (with
phases $\phi _i$), arranged in a random three-dimensional (3D) lattice
with coordinates $\vec r_i=(x_i,y_i,z_i)$. The grains are separated by
insulating boundaries producing Josephson
coupling with energy $J_{ij}=J$. The system is under the influence
of a frustrating applied magnetic field $\vec H $.

Since the field-induced effects considered in the present paper are expected
to manifest themselves in high enough applied magnetic fields (with a nearly
homogeneous distribution of magnetic flux along the junctions) and as long as
the Josephson penetration length $\lambda _J$ exceeds the characteristic size
of the Josephson network $d$ (which is related to the projected junction area
$S$, where the field penetration actually occurs, as follows $S=\pi d^2$),
the Josephson current-induced "self-field" effects (which are important for
a large-size junctions and/or small applied magnetic fields) may be safely
neglected~\cite{ref11}. Besides, it is known~\cite{ref12} that in discrete
JJAs pinning (by a single junction) actually concurs with the "self-field"
effects. Specifically, it was found~\cite{ref12} that the ratio
$d/\lambda _J$ is related to the dimensionless pinning strength parameter
$\beta$ as $d/\lambda _J=\sqrt{\beta}$ suggesting that a weak pinning regime
(with $\beta \ll 1$) simultaneously implies a smallness of the "self-field"
effect and vice versa.
And since artificially prepared Josephson networks allow for more flexibility
in varying the experimentally-controlled parameters, it is always possible
to keep the both above effects down by appropriately tuning the ratio
$d/\lambda _J$. Typically~\cite{ref7,ref8,ref9}, in this kind of experiments
$S=0.01-0.1\mu m^2$ and $d\ll \lambda _J$.

In what follows, we are interested in the magnetic field induced behavior
of the electric polarization in a $3D$ JJA at zero temperature.
Recall that a conventional (zero-field) pair polarization operator within
the model under discussion reads~\cite{ref13}
\begin{equation}
\vec p=\sum_{i=1}^Nq_i \vec r_i,
\end{equation}
where $q_i =-2en_i$ with $n_i$ the pair number operator, and $r_i$ is the
coordinate of the center of the grain.

In view of Eqs.(1)-(6), and taking into account a usual "phase-number"
commutation relation, $[\phi _i,n_j]=i\delta _{ij}$, the evolution of the
pair polarization operator is determined via the equation of motion
\begin{equation}
\frac{d\vec p}{dt}=\frac{1}{i\hbar}\left[ \vec p,{\cal H}\right ]
=\frac{2e}{\hbar }\sum_{ij}^NJ\sin \phi _{ij}(\vec H)\vec r_{ij}
\end{equation}
Resolving the above equation, we arrive at the following net value of the
magnetic-field induced polarization (per grain)
\begin{eqnarray}
\vec P(\vec H)&\equiv &\frac{1}{N}\overline {<\vec p(t)>}\nonumber\\
&=&\frac{2eJ}
{\hbar \tau N} \int\limits_{0}^ {\tau }dt \int
\limits_{0}^{t}dt'\sum_{ij}^N <\sin \phi _{ij}(\vec H)\vec r_{ij}>,
\end{eqnarray}
where $<...>$ denotes a configurational averaging over the grain positions,
while the bar means a temporal averaging (with a characteristic time $\tau$).
To consider a field-induced polarization only, we assume that
in a zero magnetic field, $\vec P\equiv 0$, implying $\phi _{ij}(0)\equiv0$.

To obtain an explicit form of the field dependence of polarization, let us
consider a site-type positional disorder allowing for weak displacements of
the grain sites from their positions of the original $3D$ lattice,
i.e., within a radius $d$ (which is of the order of the Josephson
lattice parameter and the characteristic junction size, see above) the new
position is chosen randomly according to the normalized distribution function
$f(\vec r, \vec R)$. It can be shown
that a particular choice of the distribution function will not change the
main qualitative results of the present
paper. So, assuming, for simplicity, the following distribution law
\begin{equation}
f(\vec r, \vec R)=f_r(\vec r)f_R(\vec R),
\end{equation}
with
\begin{equation}
f_r(\vec r)=\frac{1}{(2\pi d^2)^{3/2}}e^{-(x^2+y^2+z^2)/2d^2},
\end{equation}
and
\begin{equation}
f_R(\vec R)=\delta (X-d)\delta (Y-d)\delta (Z-d),
\end{equation}
we observe that the magnetic field $\vec H=(0,0,H_z)$ (applied along the
$z$-axis) will induce a non-vanishing longitudinal (along $x$-axis)
electric polarization
\begin{equation}
P_x(H_z)=P_0G(H_z/H_0),
\end{equation}
with
\begin{equation}
G(z)=ze^{-z^2}
\end{equation}
Here $P_0=ed\tau J/\hbar $, $H_0=\Phi _0/S$ with $S=\pi d^2$ being an
average projected area of a single junction, and $z=H_z/H_0$.

For small applied fields ($z\ll 1$), the induced polarization
$P_x(H_z)\approx \alpha _0H_z$, where
$\alpha _0=e\tau dJ/\hbar H_0$ is the so-called linear magnetoelectric
coefficient~\cite{ref14}. However, as we mentioned in the very beginning,
to correctly describe any induced effects in very small external fields,
both Josephson junction pinning and "self-field" effects have to be taken
into account~\cite{ref12}.
\vspace*{-0.1cm}
\begin{figure}
\epsfxsize=8.0cm
\centerline{\epsffile{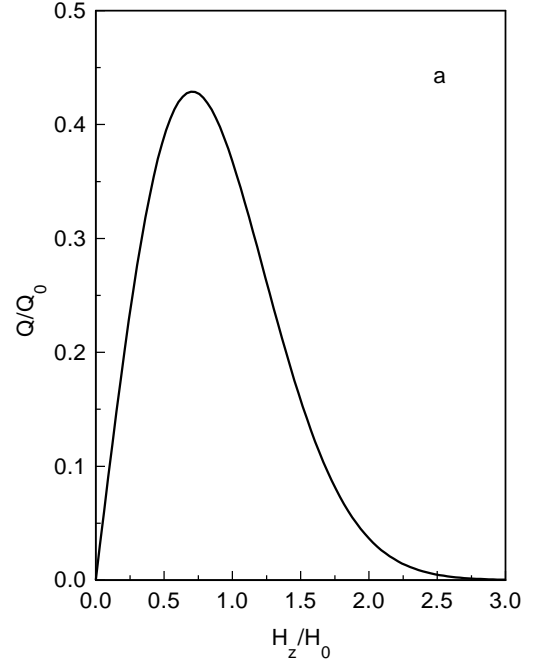} }
\vspace*{-0.1cm}
\epsfxsize=8.0cm
\centerline{\epsffile{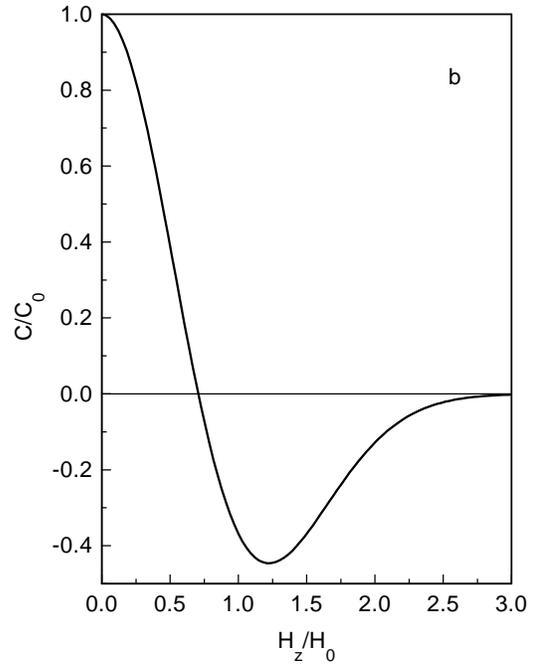} }
\vglue -0.2cm
\caption{The behavior of the induced effective charge $Q/Q_0$ (a) and
the flux capacitance $C/C_0$ (b) in applied magnetic field $H_z/H_0$.}
\end{figure}
At the same time, in view of Eq.(6) the induced polarization is related
to the corresponding change of the effective charge $Q\equiv (1/d)
\sum_i<q_ix_i>$ in applied magnetic field as follows
\begin{equation}
Q(H_z)=\frac{P_x(H_z)}{d}=Q_0G(H_z/H_0),
\end{equation}
where $Q_0=e\tau J/\hbar$.

It is of interest also to consider the related field behavior of the
effective flux capacitance $C\equiv \tau dQ(H_z)/d\Phi $ which in view of
Eq.(14) reads
\begin{equation}
C(H_z)=C_0\left( 1-2\frac{H_z^2}{H_0^2}\right )e^{-H_z^2/H_0^2},
\end{equation}
where $\Phi =SH_z$, and $C_0=\tau Q_0/\Phi _0$.
Figure 1 shows the behavior of the induced effective charge $Q/Q_0$ (a) and
the corresponding capacitance $C/C_0$ (b) in applied magnetic field $H_z/H_0$.
As is seen, at $H_z/H_0\approx 0.8$ the effective charge reaches its maximum
while the capacitance changes its sign at this field, suggesting a
significant redistribution of the junction charge balance in a model system
under the influence of an applied magnetic field, near the Josephson critical
field $H_0$. Note that a somewhat similar behavior of the magnetic field
induced charge (and related capacitance) has been observed in 2D electron
systems~\cite{ref15}.

Taking $\tau =10^{-10}s$ for the Josephson relaxation time (which is
related to the Josephson plasma frequency, $\omega _p\simeq \tau ^{-1}$,
known to be the characteristic frequency of the system for
$E_c\ll J$ regime, with $E_c$ being the Coulomb grain's charge energy;
typically~\cite{ref7} $\omega _p=10^9-10^{11}Hz$) and $J/k_B=90K$ for
a zero-temperature Josephson energy in $YBCO$ ceramics, we arrive at the
following estimates of the effective charge $Q_0\approx 10^{-16}C$, flux
capacitance $C_0\approx 10^{-11}F$, the equivalent current $I_0\sim Q_0/\tau
\approx 10^{-6}A$, and voltage $V_0\sim Q_0/C_0\approx 10^{-5}V$.
We note that the above set of estimates fall into the range of parameters
used in typical experiments to study the charging effects both in single
JJs (with the working frequency from RF range of $\omega \simeq 10 GHz$
used to stimulate the system~\cite{ref9}) and JJAs~\cite{ref10} suggesting
thus quite an optimistic possibility to observe the above-predicted field
induced effects experimentally, using a specially prepared system of arrays
of superconducting grains.

In summary, a zero-temperature behavior of the induced electric polarization,
effective charge and concomitant flux capacitance in an applied magnetic
field have been considered within a $3D$ model of Josephson junction arrays.
A possibility of the experimental observation of the discussed effects in
artificially prepared system of superconducting junctions was suggested.

\acknowledgments

The stimulating discussions with Jorge Jos\'e (Northeastern University,
Boston) are highly appreciated.
This work was financially supported by the Brazilian agency CNPq.

\end{document}